\documentstyle[twocolumn,aps,epsfig]{revtex}
\newcommand{\be}{\begin{equation}}
\newcommand{\ee}{\end{equation}}
\newcommand{\bea}{\begin{eqnarray}}
\newcommand{\eea}{\end{eqnarray}}
\renewcommand{\d}{{{\rm d}}}

\newcommand{\lton}{\mathrel{\lower.9ex
                  \hbox{$\stackrel{\displaystyle <}{\sim}$}}}

\begin{document}
\title{Kaon Interferometry: 
A Sensitive Probe of the QCD Equation of State?}
\author{Sven~Soff$^1$, Steffen~A.~Bass$^{2,3}$, David~H.~Hardtke$^1$, 
and Sergey~Y.~Panitkin$^4$}
\address{$^1$Nuclear Science Division 70-319, 
Lawrence Berkeley National Laboratory, 
1 Cyclotron Road, Berkeley, CA94720, USA\\
$^2$Department of Physics, Duke University, Durham, NC27708, USA\\
$^3$RIKEN BNL Research Center, Brookhaven National Laboratory, 
Upton, NY11973, USA\\
$^4$Physics Department, Brookhaven National Laboratory, PO Box 5000,
Upton, NY11973, USA} 
\date{\today}
\maketitle   
\vspace*{-1.0cm}
\begin{abstract}
We calculate the kaon interferometry radius parameters 
for high energy heavy ion collisions, assuming a first order
phase transition from a thermalized Quark-Gluon-Plasma (QGP) 
to a gas of hadrons. 
At high transverse momenta $K_T\sim 1\,$GeV/c direct emission 
from the phase boundary becomes important, the emission duration signal, i.e.,
the $R_{\rm out}/R_{\rm side}$ ratio, and its 
sensitivity to $T_c$ (and thus to the latent heat) are enlarged.
The QGP+hadronic rescattering transport model calculations 
do not yield unusually large radii ($R_i \le 9\,$fm). 
Finite momentum resolution effects have a strong impact 
on the extracted interferometry parameters ($R_i$ and $\lambda$) as well as on  
the ratio $R_{\rm out}/R_{\rm side}$. 
\end{abstract}
\narrowtext
\vspace*{-1.3cm}
In this Letter we present predictions for 
the kaon interferometry measurements in Au+Au collisions at the 
Relativistic Heavy Ion Collider (RHIC) 
(nucleon-nucleon center-of-mass energies up to $\sqrt{s}_{NN}=200\,$GeV). 
Correlations of identical kaon pairs as well as 
Bose-Einstein correlations in general represent an important tool 
for the understanding of the space-time dynamics in 
multiparticle production processes~\cite{Goldhaber}. 
In the particular case of relativistic heavy ion collisions, 
one important goal is to prove  the existence 
of a phase transition from quark-gluon matter 
to hadrons as predicted by QCD lattice calculations for high temperatures. 
Moreover, 
the properties of that 
phase transition, for example the critical temperature $T_c$ or its  
latent heat, are of great interest.    
For a first order phase transition, the associated 
large hadronization time was predicted to lead to 
unusually large interferometry radii~\cite{pratt86,schlei,dirk1}. 
The mixed phase of coexisting hadrons and partons 
prolongs the emission duration of particles from the 
phase boundary because of the large latent heat. This 
should then lead to  
an increase of the effective source size, in particular 
in the {\it outward} direction, i.e., parallel to the 
transverse pair velocity. 
This phenomenon was also expected to depend 
on the critical temperature $T_c$, the latent heat or 
the specific entropy of the quark-gluon phase  
(for recent reviews, see, e.g., \cite{reviews,wiedemannrep}).
However, as demonstrated recently \cite{soffbassdumi},
the subsequent hadronic rescattering phase following
hadronization from a thermalized QGP does not only
modify the pion interferometry radii but even dominates them.
Late numerous soft collisions in the hadronic phase 
diminish and also alter qualitatively
the particular dependencies on the QGP-properties~\cite{soffbassdumi}.

Here, we  investigate what can be learned from the 
interferometry analysis of kaons. 
The motivation to extend the interferometry analysis from pions to kaons 
(which are measured soon at RHIC)
is provided by several aspects.      
(i) Kaons are expected to be less contaminated by resonance decays 
compared to pions \cite{Gyulassy:1989yr,Sullivan:wb}. 
(ii) In the case of neutral kaon correlations,  
two-particle Coulomb interactions do not distort the Bose-Einstein 
correlations (while non-negligible strong final state K$\overline{\rm K}$ 
interactions persist due to the near threshold resonances 
a0(980) and f0(980) \cite{Lednicky:1982su}).  
(iii) The kaon density is considerably smaller than the pion density. 
The pion multiplicity itself has increased by approximately 
$70\%$ from SPS ($\sqrt{s}_{NN}=17.4\,$GeV) to 
RHIC ($\sqrt{s}_{NN}=130\,$GeV) \cite{Back:2000gw} 
whereas the interferometry radii are almost the same \cite{STARpreprint,Johnson:2001zi}. 
Hence, higher multiparticle correlation effects that might 
play a role for pions, should be of minor importance for 
kaons. The comparison of kaon and pion correlation functions will 
provide a test of the presently applied two-particle 
correlation formalism.   
(iv) 
The strangeness distillation mechanism \cite{Greiner:1987tg}
might further increase the time delay signature $R_{\rm out}/R_{\rm side}$. 
Kaon evaporation could lead to 
strong temporal emission asymmetries between kaons and antikaons 
\cite{Soff:1997nb}, thus probing  
the latent heat of the phase transition. 

We will show in the following that for kaons, as for the pions,  
the bulk properties  of the two-particle correlation functions 
are dominated by a long-lived hadronic rescattering phase. 
Thus, the interferometry radii appear to depend only weakly on the 
precise properties of the QGP, such as the 
thermalization time $\tau_i$, $T_c$, the latent heat or the 
specific entropy of the QGP. 
However, we will demonstrate that this sensitivity is 
considerably enlarged at high transverse momenta $K_T\sim 1\,$GeV/c. 
In this kinematic regime, direct emission from 
the phase boundary becomes important ($\sim 30\%$) 
allowing us to   
inspect the prehadronic phase with less distortions.      
Finally, we will calculate explicitly the correlation 
functions and extract the corresponding parameters with and without 
taking finite momentum resolution effects into account. 
A strong impact on the radii, the $\lambda$ intercept parameter 
and the $R_{\rm out}/R_{\rm side}$ ratio are 
observed; they all decrease, depending on $K_T$, 
if a finite momentum resolution is taken into account. 

The calculations are performed within the framework of
a relativistic transport model that describes the initial dense phase
of a QGP by means of ideal hydrodynamics employing a bag model 
equation of state that exhibits a first order phase
transition~\cite{dirk1,gersdorff}. 
Hence, we focus on a phase
transition in local equilibrium, proceeding through the formation of a mixed
phase. 
Smaller radii and emission times may result for a crossover
\cite{dirk1,Zschiesche:2001dx} or for a   
rapid out-of-equilibrium phase transition
similar to spinodal decomposition~\cite{spino}.
Cylindrically
symmetric transverse expansion and longitudinally boost-invariant
scaling flow are assumed \cite{dirk1,gersdorff,DumRi}.
This approximation should be reasonable
for central collisions at high energy, and around midrapidity.
The model reproduces the measured $p_T$-spectra and rapidity densities
of a variety of hadrons at $\sqrt{s}_{NN}=17.4\,$GeV (CERN-SPS energy),
when assuming the standard thermalization
(proper) time $\tau_i=1$~fm/c, and an entropy per net
baryon ratio of $s/\rho_B=45$ \cite{DumRi,hu_main}.
Due to the higher density at midrapidity,
thermalization may be faster at BNL-RHIC energies -- here
we assume $\tau_i=0.6$~fm/c and $s/\rho_B=200$.
(With these initial conditions results on the multiplicity, the
transverse energy, the $p_T$-distribution of charged hadrons, and
the $\overline{p}/p$ ratio are described quite
well~\cite{DumRi,hu_main}.) 
The later hadronic phase is modeled via microscopic
transport that allows us to calculate the 
so-called freeze-out, i.e., the time and coordinate space points
of the last strong interactions of an individual particle species,
rather than applying a freeze-out prescription as necessary
in the {\it pure} hydrodynamic approach.
Here, we employ a semi-classical transport model that treats
each particular hadronic reaction channel (formation and decay of hadronic
resonance states and $2\rightarrow n$ scattering)
{\em explicitly}~\cite{bass98}. The transition
at hadronization is performed by matching
the energy-momentum tensors and conserved currents of the
hydrodynamic solution and of the microscopic transport model, respectively
(for details, see~\cite{hu_main}).
The microscopic model propagates each individual hadron
along a classical trajectory, and performs $2\rightarrow n$ and $1\rightarrow
m$ processes stochastically.
Meson-meson and meson-baryon cross sections are modeled via resonance 
excitation and also contain an elastic contribution. 
All resonance properties are taken from~\cite{PDG}.
The $\pi K$ cross section for example is either elastic or is dominated 
by the $K^*(892)$,
with additional contributions from higher energy states.
In this way, a good description of elastic and total kaon cross sections
{\em in vacuum} is obtained~\cite{bass98}.
Medium effects on the hadron properties, as for example 
recently studied by hydrodynamical 
calculations employing a chiral equation of state \cite{Zschiesche:2001dx}, 
are presently neglected. 
For further details of this dynamical two-phase transport model, 
we refer to refs.~\cite{DumRi,hu_main}.

For the correlation analysis, a coordinate system is used in which
the {\it long} axis ($z$) is parallel to
the beam axis, where the {\it out} direction
is parallel to the transverse momentum vector
${\bf K_T}=({\bf p_{1T}} + {\bf p_{2T}})/2$
of the pair, and the {\it side} direction is perpendicular
to both.
Thus, $R_{\rm out}$ probes the spatial {\it and} temporal extension
of the source while $R_{\rm side}$ only probes the spatial extension.
Thus the ratio $R_{\rm out}/R_{\rm side}$ gives a measure of the
emission duration (see also eqs.(1)-(3) and discussion below).
It has been suggested that the ratio $R_{\rm out}/
R_{\rm side}$ should increase strongly once the initial
entropy density $s_i$ becomes substantially larger than that of the hadronic
gas at $T_c$~\cite{dirk1}.
The  Gaussian radius parameters 
are obtained from a saddle-point integration over 
the classical phase space distribution of the hadrons at freeze-out 
that is identified with the Wigner density of the source, 
$S(x,K)$~\cite{Podgoretsky:1983xu,wiedemannrep,Gyulassy:1989yr,Pratt:1990zq}.
\begin{eqnarray}
\label{rs}
R_{\rm side}^2({\bf K_T})&=& \langle \tilde{y}^2 \rangle ({\bf K_T})\,,\\
R_{\rm out}^2({\bf K_T})&=& \langle (\tilde{x}-\beta_t \tilde{t})^2
\rangle ({\bf K_T}) = \langle\tilde{x}^2+\beta_t^2 \tilde{t}\,^2-2
 \beta_t\tilde{x}\tilde{t}\rangle\,,\label{ro}\\
R_{\rm long}^2({\bf K_T})&=& \langle (\tilde{z}-\beta_l \tilde{t})^2
\rangle ({\bf K_T})\,,\label{rl}
\end{eqnarray}
with
$\tilde{x}^{\mu}({\bf K_T}) = x^{\mu} - \langle {x}^{\mu}\rangle({\bf K_T})$
being the space-time coordinates relative
to the momentum dependent {\it effective source centers}.
The average in (\ref{rs})-(\ref{rl}) is taken over the
emission function, i.e.
$\langle f \rangle(K)= \int d^4x f(x) S(x,K) / \int d^4x S(x,K)$ with 
$K=(E_K,{\bf K})$. 
In the {\it osl} system ${\bf \beta}=(\beta_t,0,\beta_l)$, where
${\bf \beta}={\bf K}/E_K$ and $E_K=\sqrt{m^2+{\bf K}^2}$.
Below, we cut on midrapidity kaons ($|y|<0.5$). 
In the absence of $\tilde{x}$-$\tilde{t}$ correlations, i.e.\ in particular
at small $K_T$, a large duration of emission $\Delta \tau = \surd
{\langle \tilde{t}\,^2 \rangle}$ increases $R_{\rm out}$ relative
to $R_{\rm side}$~\cite{pratt86,schlei,dirk1}. 
For strong (positive) $\tilde{x}$-$\tilde{t}$-correlations or large 
spatial anisotropies in {\it out}- and {\it side}-direction 
($\langle \tilde{y}^2\rangle > \langle \tilde{x}^2\rangle$),  
in principle $R_{\rm out} \le R_{\rm side}$ could follow.

The absolute values of the kaon radii determined by the above expressions
(\ref{rs})-(\ref{rl}) are considerably smaller than the pion radii,
especially at low $K_T$. These pion radii are larger by a factor of
$two$ at low $K_T$ ($\leq 400\,$MeV) while at higher $K_T$ the values
become similar.
This is due to the resonance source character of pions. 
Microscopic transport calculations show that
at SPS energies ($\sqrt{s}_{NN}=17.4\,$GeV) 
about $80\%$ of the pions are emitted
from various resonances \cite{soffcris98}.
This leads to a strong substructure of the
freeze-out distributions \cite{soffcris98}, e.g.\ 
strongly non-Gaussian tails. 
The ratio $R_{\rm out}/R_{\rm side}$ for kaons is shown in Fig.~1.
Due to identical freeze-out distributions 
the correlation parameters are the same for kaons and antikaons. 
The bag parameter $B$ is varied from $380$~MeV/fm$^3$ to   
$720$~MeV/fm$^3$, (i.e., the latent heat changes by $\sim4B$), 
corresponding to critical temperatures of
$T_c\simeq160$~MeV and $T_c\simeq200$~MeV,
respectively. 
A change of $T_c$ implies a variation of the longitudinal and transverse flow
profiles on the hadronization hypersurface (which is the initial condition for
the subsequent hadronic rescattering stage). 
$R_{\rm out}/R_{\rm side}$ 
is smaller for kaons than for pions at the same small $K_T$
because of their larger mass (leading to smaller velocities which 
reduce the temporal contribution to
$R_{\rm out}$ in eq.\ (\ref{ro}) ($\beta_t^2\langle \tilde{t}^2
\rangle$) \cite{Bernard:1997bq}). 
The ratios reach a value of 1.5 at $K_T\approx m_{\pi}$ 
and $K_T\approx m_{\rm K}$ for the 
rather rapidly rising pions \cite{soffbassdumi} 
and the gradually growing kaons, respectively. 
Model calculations solely based on hadronic degrees of freedom          
generally predict smaller ratios \cite{Fields:sj,Sullivan:wb,Hardtke:2000vf}.
The sensitivity of the value of $R_{\rm out}/R_{\rm side}$ 
to the critical temperature $T_c$ 
increases strongly with $K_T$.
Higher $T_c$ speeds up hadronization but on the other hand prolongs
the dissipative hadronic phase that dominates the radii. 
Moreover, in the lower $T_c$ case, direct emission and immediate
freeze-out from the phase boundary becomes important
at large $K_T$ ($\sim 1\,$GeV/c).
High-$K_T$ kaons are strongly correlated with 
early mean emission times. 
The resonance contribution for the kaons is still quite large,
decreasing  with $K_T$ from $70$ to $50\%$ for $T_c\simeq 160\,$MeV.
However, 
most of these kaons are from $K^*(892)$
decays  
with the $K^*$ having a moderate lifetime of $\tau\approx 4\,$fm/c.
Elastic scatterings prior to freeze-out contribute on the order
of $20\%$.
The direct emission from the phase boundary, i.e., the kaon did not
suffer further collisions in the hadron gas after the particle had hadronized,
increases strongly (approximately linearly with $K_T$) 
for $T_c\simeq 160\,$MeV up to $30\%$ at $K_T=1\,$GeV/c.
For the higher $T_c$ ($\simeq 200\,$MeV) hadronization is earlier. 
The hadronic phase lasts longer and the system is rather opaque
for direct emission.
This direct emission component is not present in 
{\it pure} ideal hydrodynamical calculations 
(e.g.\ \cite{Bernard:1997bq}) for which all particles, 
also at high $K_T$, are in (local) thermodynamical equilibrium. 
Thus, there is no possibility for direct emission from 
the phase boundary and escaping the hadronic phase unperturbed.  
\begin{figure}[htp]
\centerline{\hspace{.8cm}\hbox{
\epsfig{figure=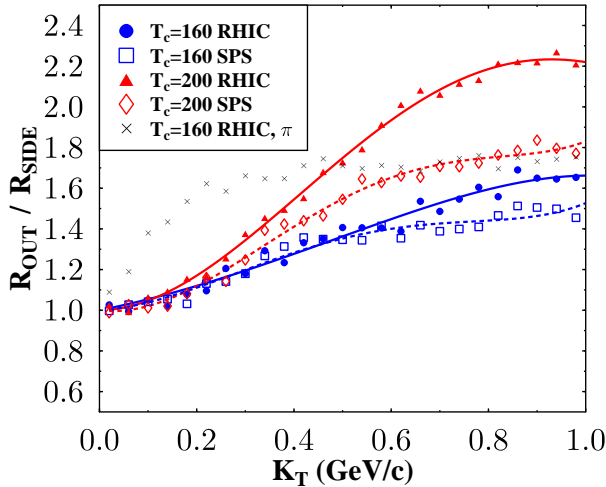,width=8cm}}}
\caption{$R_{\rm out}/R_{\rm side}$ as obtained from eqs.\ (1) and (2)
for kaons at RHIC ($\sqrt{s}_{NN}=200\,$GeV) (full symbols) and
at SPS ($\sqrt{s}_{NN}=17.4\,$GeV) (open symbols),
as a function of $K_T$ for critical temperatures
$T_c\simeq 160\,$MeV  and $T_c\simeq 200\,$MeV, respectively.
The crosses show pions for comparison. The lines are to guide the eye.}
\end{figure}
Finally, we calculate the correlation parameters (shown in Fig.\ 2) 
by performing a $\chi^2$ fit of the three-dimensional
correlation function $C_2(q_{\rm out},q_{\rm side},q_{\rm long})$
to a Gaussian as
\begin{equation}
C_2(q_{\rm o},q_{\rm s},q_{\rm l})=
1+\lambda \exp(-q_{\rm o}^2 R_{\rm o}^2-q_{\rm s}^2 R_{\rm s}^2
-q_{\rm l}^2 R_{\rm l}^2)\,.
\end{equation}
The correlation functions  are calculated from  
the phase space distributions of kaons at freeze-out
using the {\it correlation after burner} by Pratt~\cite{pratt86,Pratt:1990zq}. 
It is assumed that the particles are emitted from the large system
independently, which allows to factorize
the $N$-boson production  
amplitude into $N$ one-boson amplitudes ${\cal A}(x)$.
Then, the emission function is computed as the Wigner transform
$S(x,K) = \int \d^4y \,e^{iy\cdot K} {\cal A}^*(x+y/2) {\cal A}(x-y/2)$.
The two-boson correlation function is given by
\bea
& & C_2({\bf p},{\bf q})-1 = \nonumber\\
& & \frac
{\int\d^4x S(x,{\bf K}) \int\d^4y S(y,{\bf K}) \exp\left(
2ik\cdot(x-y)\right)}
{\int\d^4x S(x,{\bf p}) \int\d^4x S(y,{\bf q})} \nonumber\\
&\simeq& \frac
{\int\d^4x S(x,{\bf K}) \int\d^4y S(y,{\bf K}) \exp\left(
2ik\cdot(x-y)\right)}
{|\int\d^4x S(x,{\bf K})|^2}, \label{fullC}
\eea
where $2{\bf K}={\bf p}+{\bf q}$, $2{\bf k}={\bf p}-{\bf q}$, and  
$2k^0 = E_p-E_q$. The second line in~(\ref{fullC}) holds in the limit where   
the width of the correlation function is small such that
${\bf p}\sim{\bf q}\sim{\bf K}$.
\begin{figure}[htp]
\centerline{\hspace{.8cm}\hbox{
\epsfig{figure=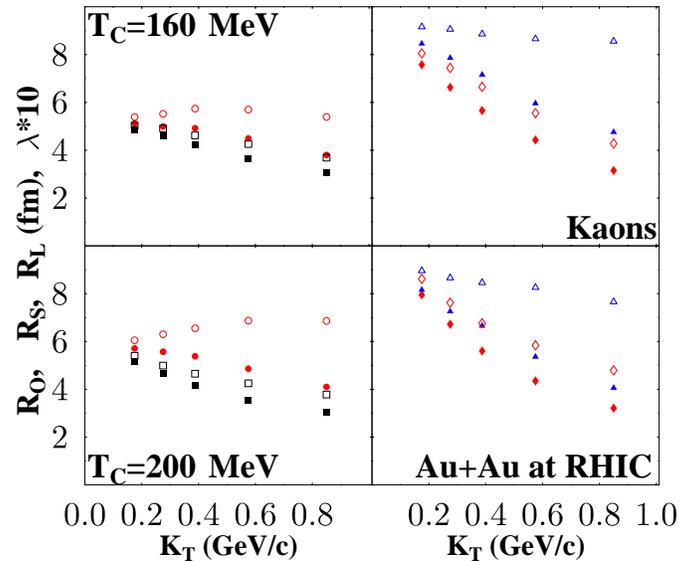,width=9cm}}}
\caption{Kaon correlation parameters 
as obtained from a $\chi^2$ fit of $C_2$ (eq.\ (5)) to the 
Gaussian {\it ansatz} (eq.\ (4)) 
for Au+Au collisions at RHIC ($\sqrt{s}_{NN}=200\,$GeV).
The top and bottom panels show the results for 
$T_c \simeq 160\,$MeV and $T_c \simeq 200\,$MeV, respectively. 
$R_{\rm out}$ (circles) and $R_{\rm side}$ (squares) are shown on the left  
while $R_{\rm long}$ (diamonds) and $\lambda \cdot 10$ (triangles) 
are shown on the right.
Full and open symbols correspond to calculations
with and without taking momentum resolution effects into account,
respectively.}
\end{figure}
Given a model for a chaotic source described by $S(x,{\bf K})$, such as
the transport model described above, eq.~(\ref{fullC}) can be employed to
compute the correlation function.
The expressions (\ref{rs})-(\ref{rl}), based on an Gaussian {\it ansatz},  
yield larger values for the pion radii 
than performing a fit to the correlation functions.  
For the kaon transverse radii, similar results 
are obtained with both methods.  
Only $R_{\rm long}$ is larger if determined 
by (\ref{rl}) as compared to the fitting result. 
This is due to the non-Gaussian contribution arising from 
the longitudinal expansion dynamics that is similar for pions 
and kaons \cite{Hardtke:2000vf}. 
Kaons are better candidates for the Gaussian expressions.   
There are fewer resonance decays into kaons compared 
to pions, and long-lived resonances do not play as important a role.

For $T_c \simeq 200\,$MeV, $R_{\rm out}$ is only approximately $1\,$fm larger
than in the $T_c \simeq 160\,$MeV case. This reflects a fact already known from
the pions. Higher $T_c$ leads to an earlier hadronization, thus
causing a prolonged hadronic phase.
When taking finite momentum resolution ({\it f.m.r.}) into account, 
the {\it true} particle momentum $p$
obtains an additional random component. This random component is
assumed to be Gaussian with a width $\delta p$.
The relative momenta of pairs are then calculated from these modified momenta.
However, the correlator is calculated with the {\it true} relative
momentum.
While $R_{\rm out}$ remains constant or even
slightly increases with $K_T$
when calculated without {\it f.m.r.}, it drops if
a {\it f.m.r.\ }of $\approx 2\%$ of the center of 
each $K_T$ bin is considered,
a value assumed for the STAR detector \cite{STARpreprint}.
Accordingly, the differences with and without {\it f.m.r.\ }increase 
with $K_T$.
The {\it f.m.r.\ }leads to smaller radii.
$R_{\rm out}$ is strongly reduced while $R_{\rm side}$
shows a moderate reduction. Thus, the $R_{\rm out}/R_{\rm side}$ ratio
is considerably reduced through the {\it f.m.r.}.
For example, in the $T_c \simeq 200\,$MeV case,
it is reduced from  $1.8$ to $1.35$. However, it is always larger than one.
A proper correction for f.m.r.\ 
(STAR and other experiments \cite{na44prlnew} do correct)
is needed to maintain the sensitivity to the QGP-properties.
The $\lambda$ parameter is roughly constant as function of
$K_T$ for $\delta p /p =0$ but it decreases rapidly 
with a {\it f.m.r.}.
The {\it correlation strength is transported to larger $q$ values} 
by the {\it f.m.r.\ }effects.

We have calculated kaon correlation parameters 
for Au+Au collisions at RHIC energies, assuming a first-order phase 
transition from a thermalized QGP to a gas of hadrons. 
No unusually large radii are seen ($R_i\le 9\,$fm).  
A strong direct emission component from the phase boundary is found at 
high transverse momenta ($K_T\sim 1\,$GeV/c) where also the 
sensitivity to the critical temperature, the latent heat and 
specific entropy of the QGP is enlarged. 
Finite momentum resolution effects reduce the {\it true} 
parameters ($R_i,\,\lambda$) and the ratio $R_{\rm out}/R_{\rm side}$ 
substantially. 
Kaon results from RHIC at high $K_T$ will provide 
an excellent probe of the space-time dynamics close to the 
phase-boundary and to the properties of this 
prehadronic state, possibly an equilibrated Quark-Gluon-Plasma.
\vspace*{-0.6cm}
\acknowledgements 
\vspace*{-0.5cm}
We are grateful to A. Dumitru, 
M.\ Gyulassy, M.\ Lisa, L.\ McLerran, 
S.\ Pratt, D.H.\ Rischke, R.\ Snellings, H. St\"ocker,
X.N.\ Wang, and N.\ Xu for many valuable comments. We thank the
UrQMD collaboration for permission to use the UrQMD transport model 
and S.\ Pratt for providing the correlation program CRAB. 
S.S.\ has been supported by the Alexander von Humboldt Foundation and 
DOE Grant No.\ DE-AC03-76SF00098.
S.A.B.\ acknowledges support from DOE Grant No.\ DE-FG02-96ER40945 
and DE-AC02-98CH10886.


\vspace*{-0.6cm}
\begin{references}
\vspace*{-1.6cm}
\bibitem{Goldhaber}
G.~Goldhaber {\it et al.,}
Phys.\ Rev.\  {\bf 120}, 300 (1960);
G.~I.~Kopylov, M.~I.~Podgoretsky,
Sov.\ J.\ Nucl.\ Phys.\  {\bf 15}, 219 (1972);
E.~Shuryak,
Phys.\ Lett.\ B {\bf 44}, 387 (1973);
G.~I.~Kopylov,
Phys.\ Lett.\ B {\bf 50}, 472 (1974); 
M.~Gyulassy {\it et al.,} 
Phys.\ Rev.\ C {\bf 20}, 2267 (1979);
A.\ Makhlin, Y.\ Sinyukov,
Z.\ Phys.\ C {\bf 39}, 69 (1988);
Y.~Hama, S.~Padula,
Phys.\ Rev.\ D {\bf 37}, 3237 (1988);
M.~I.~Podgoretsky,
Sov.\ J.\ Part.\ Nucl.\ {\bf 20}, 266 (1989).

\bibitem{pratt86}
S.~Pratt,
Phys.\ Rev.\  D {\bf 33}, 1314 (1986);
G.~Bertsch, M.~Gong, M.~Tohyama, Phys.\ Rev.\ C {\bf 37}, 1896 (1988).

\bibitem{schlei}
B.~R.~Schlei {\it et al.,}  
Phys.\ Lett.\  B {\bf 293}, 275 (1992);
J.~Bolz {\it et al.,} 
Phys.\ Rev.\  D {\bf 47}, 3860 (1993).

\bibitem{dirk1}
D.\ Rischke, M.~Gyulassy, Nucl.~Phys.~{\bf A608}, 479 (1996).


\bibitem{reviews}
R.~M.~Weiner,
Phys.\ Rept.\ {\bf 327}, 249 (2000);
T.~Cs\"org\H o,
hep-ph/0001233; 
G.~Baym,
Acta Phys.\ Polon.\ B {\bf 29}, 1839 (1998).


\bibitem{wiedemannrep}
U.\ Wiedemann, U.~Heinz,
Phys.\ Rept.\  {\bf 319}, 145 (1999).


\bibitem{soffbassdumi}
S.~Soff {\it et al.},
Phys.\ Rev.\ Lett.\  {\bf 86}, 3981 (2001).

\bibitem{Gyulassy:1989yr}
M.~Gyulassy, S.\ Padula, 
Phys.\ Rev.\ C {\bf 41}, 21 (1990);
Phys.\ Lett.\  B {\bf 217}, 181 (1989).

\bibitem{Sullivan:wb}
J.~P.~Sullivan {\it et al.},
Phys.\ Rev.\ Lett.\  {\bf 70}, 3000 (1993).



\bibitem{Lednicky:1982su}
R.~Lednicky, V.~L.~Lyuboshits,
Sov.\ J.\ Nucl.\ Phys.\  {\bf 35}, 770 (1982);
Y.~Sinyukov {\it et al.}, 
Phys.\ Lett.\ B {\bf 432}, 248 (1998).

\bibitem{Back:2000gw}
PHOBOS Collaboration, B.~B.~Back {\it et al.},
Phys.\ Rev.\ Lett.\  {\bf 85}, 3100 (2000);
nucl-ex/0108009.

\bibitem{STARpreprint}
STAR Collaboration, C.~Adler {\it et al.},
Phys.\ Rev.\ Lett.\  {\bf 87}, 082301 (2001); 
S.~Panitkin,
nucl-ex/0106018.


\bibitem{Johnson:2001zi}
PHENIX Collaboration, S.~C.~Johnson {\it et al.},
Nucl.\ Phys.\ {\bf A698}, 603 (2002);
W.~A.~Zajc {\it et al.},
nucl-ex/0106001.


\bibitem{Greiner:1987tg}
C.~Greiner {\it et al.,} 
Phys.\ Rev.\ Lett.\  {\bf 58}, 1825 (1987);
C.~Spieles {\it et al.},  
Phys.\ Rev.\ Lett.\  {\bf 76}, 1776 (1996).


\bibitem{Soff:1997nb}
S.~Soff {\it et al.},
J.\ Phys.\ G {\bf 23}, 2095 (1997); 
D.~Ardouin {\it et al.},
Phys.\ Lett.\ B {\bf 446}, 191 (1999).


\bibitem{gersdorff}
H.~Von Gersdorff {\it et al.},
Phys.\ Rev.\ D {\bf 34}, 2755 (1986).

\bibitem{Zschiesche:2001dx}
D.~Zschiesche {\it et al.}, 
nucl-th/0107037.

\bibitem{spino}
T.~Cs\"org\H o,  L.P.~Csernai,
Phys.\ Lett.\ B {\bf 333}, 494 (1994);
H.~Heiselberg, A.~D.~Jackson,
nucl-th/9809013;
J.~Rafelski, J.~Letessier,
Phys.\ Rev.\ Lett.\  {\bf 85}, 4695 (2000); 
A.~Dumitru, R.~Pisarski, Phys.\ Lett.\ B {\bf 504}, 282 (2001); 
O.~Scavenius {\it et al.},
Phys.\ Rev.\ Lett.\  {\bf 87}, 182302 (2001).


\bibitem{DumRi}
A.~Dumitru and D.\ Rischke,
Phys.\ Rev.\  C {\bf 59}, 354 (1999).

\bibitem{hu_main}
S.~A.~Bass, A.~Dumitru, Phys.\ Rev.\  C {\bf 61}, 064909 (2000).


\bibitem{bass98}
S.~A.~Bass {\it et al.}, 
Prog.~Part.~Nucl.~Phys.~{\bf 41}, 255 (1998);
M.~Bleicher {\it et al.}, J.\ Phys.\ G {\bf 25}, 1859 (1999).

\bibitem{PDG}
Review~of~Particle~Physics,
Eur.~Phys.~J.~C~{\bf 3}, 1 (1998).

\bibitem{Pratt:1990zq} 
S.~Pratt {\it et al.,} 
Phys.\ Rev.\ C {\bf 42}, 2646 (1990); 
S.~Pratt,
Phys.\ Rev.\ Lett.\ {\bf 53}, 1219 (1984);
Phys.\ Rev.\  C {\bf 49}, 2722 (1994);
W.~A.~Zajc, Phys.\ Rev.\ D {\bf 35}, 3396 (1987);
S.\ Pratt {\it et al.}, Nucl.\ Phys.\ {\bf A566}, 103c (1994).

\bibitem{Podgoretsky:1983xu}
M.~I.~Podgoretsky,
Sov.\ J.\ Nucl.\ Phys.\  {\bf 37}, 272 (1983).


\bibitem{soffcris98}
S.~Soff {\it et al.},
CRIS'98, Acicastello, Italy, 1998;
proceedings {\it Measuring the size of things in the universe:
HBT interferometry and heavy ion physics}; edited
by S.\ Costa {\it et al.}, 
World Scientific, (1999), 221-233.


\bibitem{Bernard:1997bq}
S.~Bernard {\it et al.}, 
Nucl.\ Phys.\ A {\bf 625}, 473 (1997).

\bibitem{Fields:sj}
D.~E.~Fields {\it et al.},
Phys.\ Rev.\ C {\bf 52}, 986 (1995).


\bibitem{Hardtke:2000vf}
D.~Hardtke, S.~Voloshin,
Phys.\ Rev.\ C {\bf 61}, 024905 (2000).

\bibitem{na44prlnew}
NA44 Collaboration, I.\ G.\ Bearden {\it et al.}, 
Phys.\ Rev.\ Lett.\ {\bf 87}, 112301 (2001).

\end{references}
\end{document}